\documentclass[preprint,12pt]{elsarticle}

\usepackage{lineno,hyperref}
\modulolinenumbers[5]

\newcounter{bla}

\journal{Computer Physics Communications}

\usepackage{amsfonts}
\usepackage{amsmath}
\usepackage{amssymb}
\usepackage{booktabs}
\usepackage{tabularx}
\usepackage{mathtools}
\usepackage{pgfplots}
\usepackage{listings}
\usepackage{geometry}
\usepackage{pdflscape}
\usepackage{enumerate}
\usepackage{esvect}
\usepackage[figuresright]{rotating}
\usepackage{footnote}
\usepackage{listings}
\usepackage{xcolor}
\usepackage{enumitem}
\usepackage{tocloft}
\usepackage{textcomp}
\usepackage{braket}
\usepackage{nicefrac}
\usepackage{nccmath}
\usepackage{bm}
\usepackage{import}
\usepackage{dpfloat}
\usepackage{nicefrac}
\usepackage{subcaption}
\usepackage{float}
\usepackage{afterpage}
\usepackage{multirow}
\usepackage{siunitx}
\usepackage{graphicx}
\usepackage{tikz}
\definecolor{Darkgreen}{rgb}{0,0.4,0}
\usepackage[ruled]{algorithm2e}
\graphicspath{{Figs/}}
\DeclarePairedDelimiter\abs{\lvert}{\rvert}%
\definecolor{light-gray}{gray}{0.95}
\newcommand{\code}[1]{\colorbox{light-gray}{\texttt{#1}}}
\usepackage{graphicx}
\newcommand{\CC}{%
    {\settoheight{\dimen0}{C}C\kern-.05em \resizebox{!}{\dimen0}{\raisebox{\depth}{++}}}}
\newcommand{\CCC}{%
    {\settoheight{\dimen0}{C}C/C\kern-.05em \resizebox{!}{\dimen0}{\raisebox{\depth}{++}}}}
\newcommand{\CS}{%
    {\settoheight{\dimen0}{C}C\kern-.05em \resizebox{!}{\dimen0}{\raisebox{\depth}{\#}}}}

\definecolor{mygray}{rgb}{0.4,0.4,0.4}
\definecolor{mygreen}{rgb}{0,0.8,0.6}
\definecolor{myorange}{rgb}{1.0,0.4,0}

\begin{document}
    
    \begin{frontmatter}
        
        \title{HoloGen: An open-source toolbox for high-speed hologram generation}
                
        \author[mymainaddress]{Peter J. Christopher\corref{mycorrespondingauthor}}
        \cortext[mycorrespondingauthor]{Corresponding author}
        \ead{pjc209@cam.ac.uk}
        \ead[url]{www.peterjchristopher.me.uk}
        
        \author{Andrew Kadis}
        
        \author{George S. D. Gordon}
        
        \author{Timothy D. Wilkinson}
        
        \address[mymainaddress]{Centre of Molecular Materials, Photonics and Electronics, University of Cambridge}
    
        \begin{abstract}
            The rise of virtual and augmented reality systems has prompted an increase in interest in the fields of 2D and 3D computer-generated holography (CGH). The numerical processing required to generate a hologram is high and requires significant domain expertise. This has historically slowed the adoption of CGH in emerging fields.
            
            In this paper we introduce HoloGen, an open-source Cuda C and \CC{} framework for computer-generated holography. HoloGen unites, for the first time, a wide array of existing hologram generation algorithms with state of the art performance while attempting to remain intuitive and easy to use. This is enabled by a \CS{} and Windows Presentation Framework (WPF) graphical user interface (GUI). A novel reflection based parameter hierarchy is used to ensure ease of modification. Extensive use of \CC{} templates based on the Standard Template Library (STL), compile time flexibility is preserved while maintaining runtime performance.
            
            The current release of HoloGen unites implementations of well known generation algorithms including Gerchberg-Saxton (GS), Liu-Taghizadeh (LT), direct search (DS), simulated annealing (SA) and one-step phase-retrieval (OSPR) with less known specialist variants including weighted GS and Adaptive OSPR.
            
            Benchmarking results are presented for several key algorithms. The software is freely available under an MIT license.
        \end{abstract}
    
    
    \end{frontmatter}

    {\bf PROGRAM SUMMARY}
    
    \begin{small}
        \noindent
        {\em Program Title:} HoloGen v2.2.1.17177    \\
        {\em Licensing provisions: } MIT      \\
        {\em Programming language:} Cuda, \CCC, \CS \\
        {\em Program obtainable from:} \url{https://gitlab.com/CMMPEOpenAccess/HoloGen}   \\
        {\em Maintainers:} Peter J. Christopher, the Centre of Molecular Materials, Photonics and Electronics, University of Cambridge    \\
        {\em No. of lines in distributed program:} 76,295    \\
        {\em Distribution format:} ClickOnce installer, GitLab repository    \\
        {\em Computer:} Variable, Nvidia graphics card required   \\
        {\em Operating System:} Windows 10 or later    \\
        {\em External packages:} Cuda, ManagedCuda, MathNet, Newtonsoft.Json, NUnit, AForge, Accord, ClosedXML, CefSharp, PdfiumViewer, Xceed, NHotkey, SharpDX, MaterialSkin, Xamarin.forms, HelixToolkit, Dragablz, LiveCharts, MahApps    \\
        {\em Nature of problem:} Hologram generation for two-dimensional Fourier and Fresnel holograms displayed on amplitude or phase modulating spatial light modulators with binary or multi-level control\\
        {\em Solution method:} Algorithmic variants including Gerchberg-Saxton, Liu-Taghizadeh, Direct Search, Simulated Annealing and One-Step Phase-Retrieval. Includes real-time reporting, batch processing and complex field manipulation \\
        {\em Restrictions:} Graphical user interface only exposes access to two-dimensional hologram generation\\
        {\em Unusual features:} Includes a novel reflection based parameter hierarchy for ease of modification \\
    \end{small}


    \section{Introduction}
    
    Computer-generated holography (CGH) has been anticipated since the middle of the 1960s \cite{waters1966holographic, brown1966decentering}. It took until the 1980s for computers to offer the required performance and computer-generated holography (CGH) to see practical use \cite{stone1988hybrid}. CGH is widely used today in fields including fibre and wavelength multiplexing, image correction, image recognition, optical tweezing and video projection \cite{crossland2000holographic, dong1996design, jesacher2004diffractive, muller1974real,georgiou2008aspects,cable200453}.
    
    The ability to use CGH has, however, remained the domain of optics experts. In this work we present HoloGen, a new open-source package for CGH. To the authors' knowledge, there is no open-source package capable of generating holograms using the wide array of modern algorithms, modulation schemes and interfaces available.
    
    We start by introducing the reader to the physics of optical holography, referring them to companion materials where necessary. We then continue to cover the common classes of algorithms used and to discuss the limitations and constraints of each. We then discuss the architecture, algorithm implementation, templatisation and reflection parameter hierarchy used within HoloGen. Finally, we demonstrate HoloGen in a real-world system and draw conclusions.
    
    \section{Background}
    
\begin{figure}[tbhp]
        \centering
        {\includegraphics[width=1.0\textwidth,page=1]{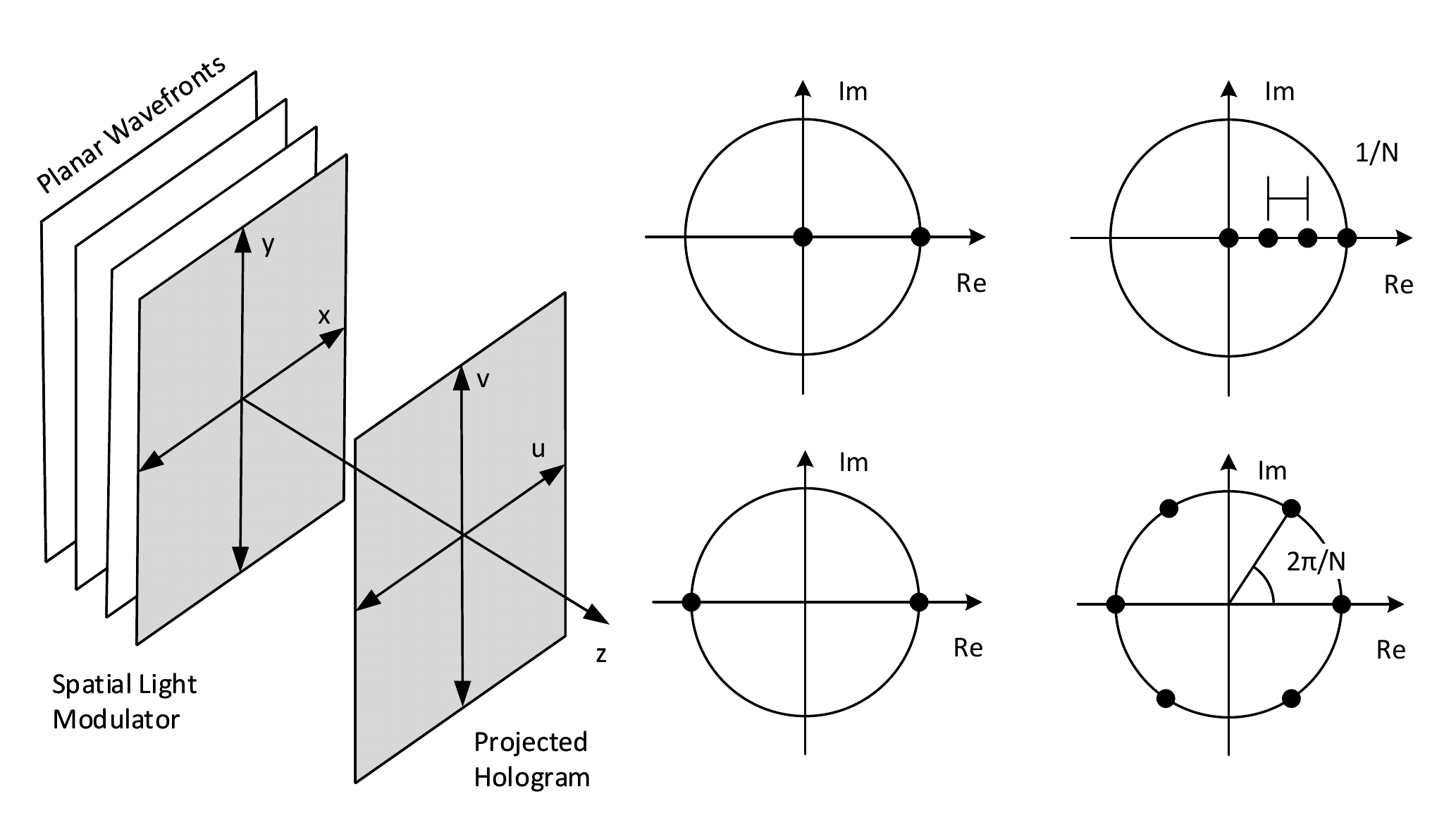}}
        \caption{Coordinate systems used in describing a hologram (left) and common spatial light modulator modulation schemes including binary amplitude (centre top), binary phase (centre bottom), multi-level amplitude (right top) and multi-level phase (right bottom)}
        \label{fig:coordsandmods}
    \end{figure}

    A spatial light modulator (SLM) with $100\%$ fill factor pixels illuminated by planar waves produces a hologram in the far-field given by a two-dimensional discrete Fourier transform (DFT) \cite{goodman2005introduction}. This is shown in Figure~\ref{fig:coordsandmods} (left).
    
    \begin{align}
    F_{u,v} = \mathcal{F}\{f_{x,y}\}         & = \frac{1}{\sqrt{N_xN_y}}\sum_{x=0}^{N_x-1}\sum_{y=0}^{N_y-1} f_{xy}e^{-2\pi i \left(\frac{u x}{N_x} + \frac{v y}{N_y}\right)} \label{fouriertrans2d5c}   \\
    f_{x,y} = \mathcal{F}^{ - 1 }\{F_{u,v}\} & = \frac{1}{\sqrt{N_xN_y}}\sum_{u=0}^{N_x-1}\sum_{v=0}^{N_y-1} F_{uv}e^{2\pi i \left(\frac{u x}{N_x} + \frac{v y}{N_y}\right)}  \label{fouriertrans2d5d},
    \end{align}
    
    where $u$ and $v$ represent the spatial frequencies and $x$ and $y$ represent the source coordinates. The fast Fourier transform (FFT) algorithm allows generation performance of $O(N_xN_y\log{N_xN_y})$ where the $x$/$u$ and $y$/$v$ respective resolutions are given by $N_x$ and $N_y$ \cite{carpenter2010graphics}. Practically this means that in order to find a given far-field hologram, we must find a discrete aperture function where $f(x,y)$ where $F(u,v) = \mathcal{F}\{f(x,y)\}$. 
    
    Fresnel or mid-field holograms are similar to their Fraunhofer or far-field counterparts save that they add an additional quadratic phase term.
    
    Real-world SLMs are capable of modulation in only a limited number of states, typically either phase or amplitude in either a binary or multi-level manner, Figure~\ref{fig:coordsandmods}. The choice between amplitude and phase modulating devices is often decided by the application and the number of modulation levels of the liquid crystal and backplane used \cite{Huang2018}. 
    
    The number of modulation levels is dependent on the technology used. Faster switching ferroelectric devices are typically binary whereas nematic devices often allow for multi-level control at the expense of switching speed.
    
    Error metrics also vary greatly by application. The phase of the replay field is unimportant in holographic projection whereas amplitude can be unimportant in applications such as optical tweezing. For human eye applications, variance is the primary concern whereas mean square error is more important in lithography. Many applications are only concerned with a portion of the replay field. Adjusting error metrics to only include these regions provides additional problem freedom for improved convergence and algorithm behaviour.
    
    \section{Algorithms}
    
    A number of algorithms are commonly used in holography. As a far-field hologram can be considered as a two-dimensional FFT, the inverse FFT (IFFT) of the target image gives the ideal SLM aperture function as a field of complex values. Real-world SLMs are capable of modulating light in only phase or amplitude, so hologram generation becomes a task of adapting the idealised aperture function to meet real-world constraints \cite{bendory2017fourier}. This process of adapting the aperture function to the real-world modulation constraints is known as \textit{quantisation}.
    
    Traditional Fourier transforms have a complexity of $O(N^4)$ for a square field of dimension $N$. An FFT of a square SLM has complexity $O(N^2 \log{N})$ \cite{cooley1965algorithm}. CGH algorithms exacerbate this, often being $O(N^2)$ themselves. For example, running simulated annealing on every pixel of a field with no further optimisations is $O(N^4 \log{N})$. Moving from a 'hd' $1080\times1920$ to '4k' $2160\times3840$ display results in a computational complexity $21$ times higher while only containing $4$ times the number of elements.
    
    The achievable quality of a hologram generation process is dependent on the \textit{three freedoms}: \textit{amplitude}, \textit{phase} and \textit{scale}. In many applications, only part of the replay field is controlled, giving amplitude freedom in the other areas. Phase freedom is often available in projection systems due to the phase insensitivity of the eye and scale freedom is available when efficiency is less important than fidelity. Exact solutions to the problem are normally impossible and compromises must be made on these three constraints in order to produce high quality holograms \cite{wyrowski1990diffractive}. 
    
    \subsection{Error Metrics}
    
    Perhaps the most common error metric used is mean squared error (MSE). The MSE $E_{MSE}(T,R)$ is given as a relationship between generated replay field $R$ to target image $T$.
    
    \begin{equation} \label{mse}
    E_{\text{MSE}}(T,R) = \frac{1}{N_x N_y}\sum_{x=0}^{x=N_x-1}\sum_{y=0}^{y=N_y-1} \left[\abs{T(x,y)} -  \abs{R(x,y)}\right]^2 
    \end{equation}
    
    When perception by the human eye is the primary goal, the structural similarity index (SSIM) often provides a more useful metric \cite{buckley2011real}
    
    \begin{equation}
    E_{SSIM}(R,R_n) = \frac{\left(2\mu_T\mu_R+c_1\right)\left(2\sigma_{TR}+c_2\right)} {\left(\mu_T^2+\mu_R^2+c_1\right)\left(\sigma_T^2+\sigma_R^2+c_2\right)}
    \end{equation}
    
    where $\sigma_R$ and $\sigma_T$ are the replay and target variances; $\mu_R$ and $\mu_T$ are the replay and target means; $\sigma_{TR}$ is the covariance of the two images; $c_1$ and $c_2$ are functions of pixel dynamic range, $L$, where $c_1=(k_1L)^2$ and $c_2=(k_2L)^2$. $k_1$ and $k_2$ are respectively usually taken as $0.01$ and $0.03$. HoloGen incorporate MSE and SSIM variants for both the phase sensitive and phase insensitive case.
    
    \subsection{Iterative Fourier Transform Algorithms} \label{ifta}
    
\begin{figure}[tbhp]
        \centering
        {\includegraphics[width=0.48\textwidth,page=1]{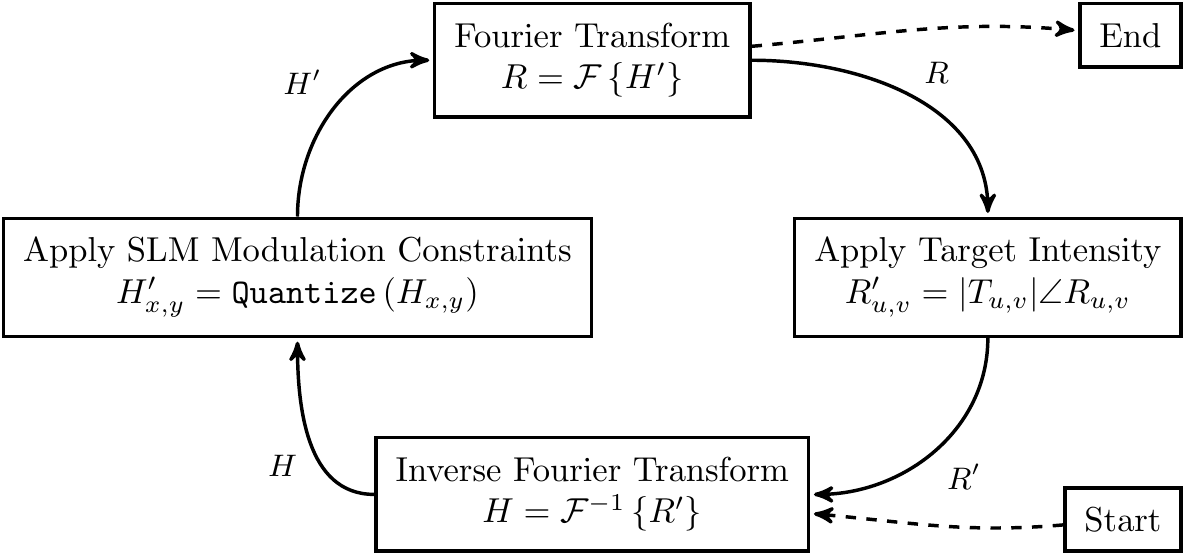}}
        {\includegraphics[width=0.48\textwidth,page=1]{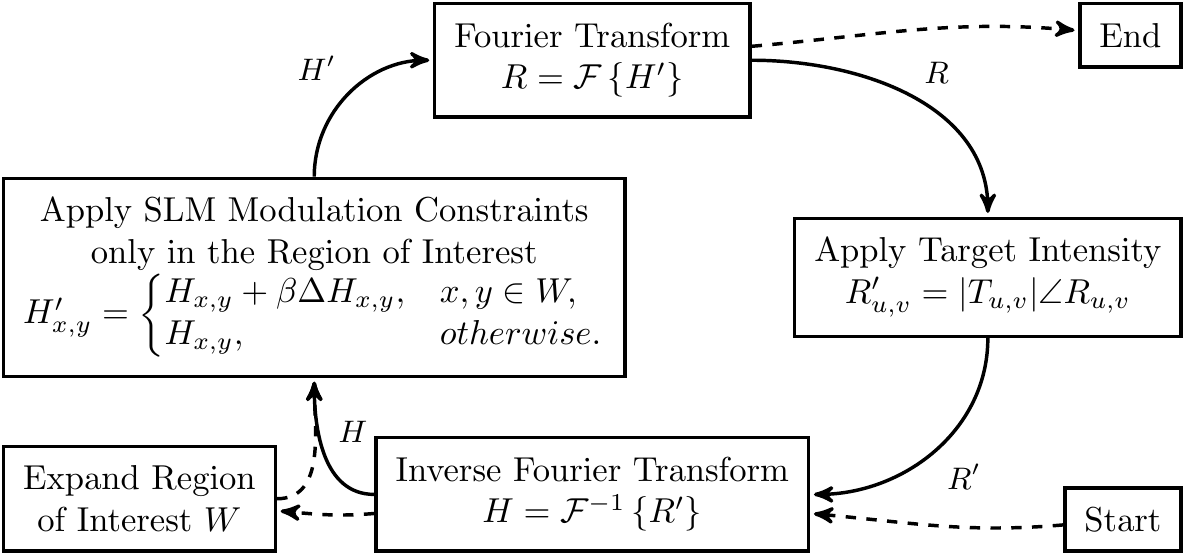}}
        \caption{Selection of iterative Fourier transform algorithms including Gerchberg-Saxton (left) and Liu-Taghizadeh (right).}
        \label{fig:algs1}
    \end{figure}

    Perhaps the most common algorithm is the Gerchberg-Saxton (GS) algorithm shown in Figure~\ref{fig:algs1} (left). Originally developed in 1972, GS was designed for phase-retrieval problems before being applied to holography \cite{gerchberg1972practical}. GS is part of the iterative Fourier transform algorithm family (IFTAs) and performs best for multi-level phase SLMs where convergence can occur in only a few iterations. \cite{fienup1979iterative}. A number of variants on GS exist which focus on target modifications, phase randomisation and weighting to improve convergence.\cite{fienup1978reconstruction}. Perhaps the most common of these is Liu-Taghizadeh (LT) shown in Figure~\ref{fig:algs1} (right) where only a portion of the replay field is initially targeted and then expanded with each iteration \cite{ liu2002iterative}.
    
    \subsection{Holographic Search Algorithms}
    
    Many SLMs offer only piecewise modulation. In these cases, approaches that rely on smooth movements such as the family of IFTA algorithms in Section~\ref{ifta} often fail to converge to an optimal solution and holographic search algorithms (HSAs) are used instead. HSAs operate by taking an initial guess at a solution and then iteratively trialling modifications to the guess. This reduces the impact of local minima.
    
    \begin{figure}[tbhp]
        \centering
        {\includegraphics[width=0.48\textwidth,page=1]{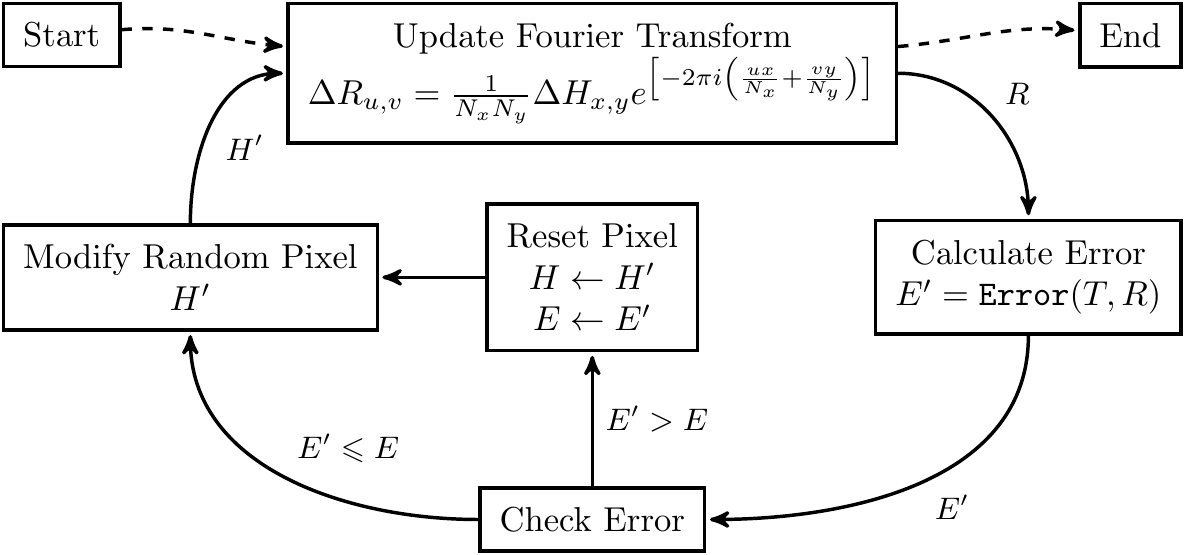}}
        {\includegraphics[width=0.48\textwidth,page=1]{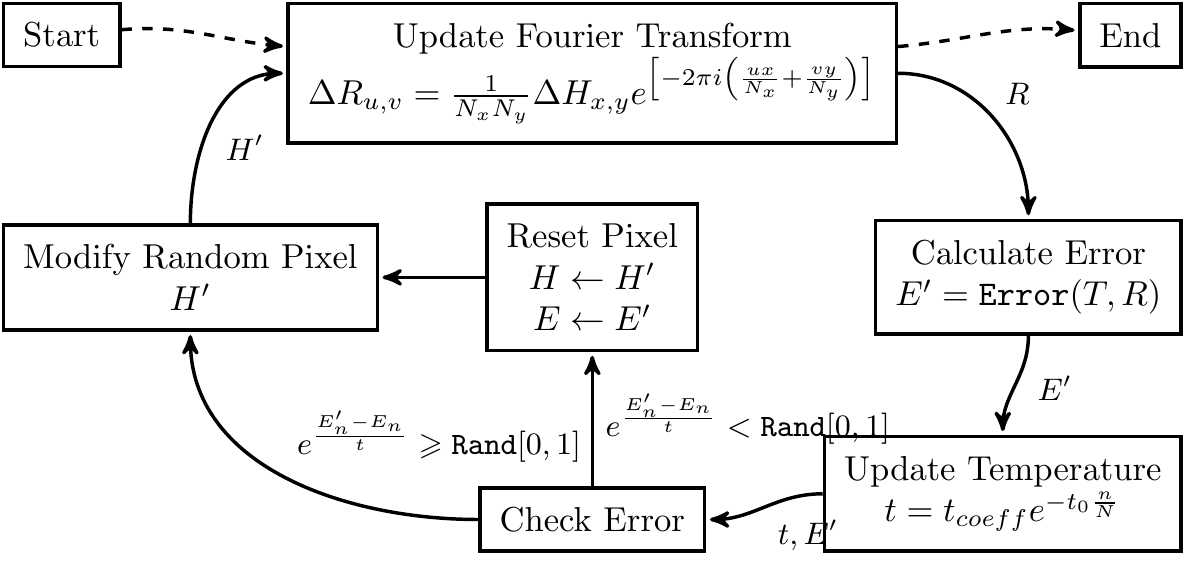}}
        \caption{Selection of holographic search algorithms including Direct Search (left) and Simulated Annealing (right).}
        \label{fig:algs2}
    \end{figure} 
    
    Perhaps most common is direct search (DS), Figure~\ref{fig:algs2} (left) \cite{jennison1989direct}. DS operates greedily, using a given error function $\texttt{Error}(T, R)$ to determine whether a given change has improved or worsened the error \cite{chhetri2000stochastic}. 
    
    Simulated annealing (SA) as shown in Figure~\ref{fig:algs2} (right), operates similarly with the addition of a temperature based probabilistic function that sometimes allows the acceptance of a worse solution \cite{kirkpatrick1983optimization}. This improves final image quality at the expense of longer run times \cite{akahori1986spectrum}.
       
    \subsection{Time-Averaged Algorithms}
    
    For low-latency or real-time display applications, a third family of techniques exists. The most well known of these is one-step phase-retrieval (OSPR), Figure~\ref{fig:algs2} (left). The time averaging effects of the human eye allow for high frame-rate SLMs to show a sequence of sub-frames in quick succession \cite{buckley200870}. The MSE of the rolling time-average of the images follows a reciprocal relationship with the number of sub-frames $N$.
    
    \begin{equation} \label{MSEdecline}
    MSE_{ospr}=\frac{1}{\sqrt{N}} \sum_{n=1}^{x=N} MSE_n
    \end{equation}
    
    The most common variant of OSPR is adaptive OSPR which uses a feedback loop to compensate for cumulative error \cite{buckley2011real}.
    
    \begin{figure}[tbhp]
        \centering
        {\includegraphics[width=0.48\textwidth,page=1]{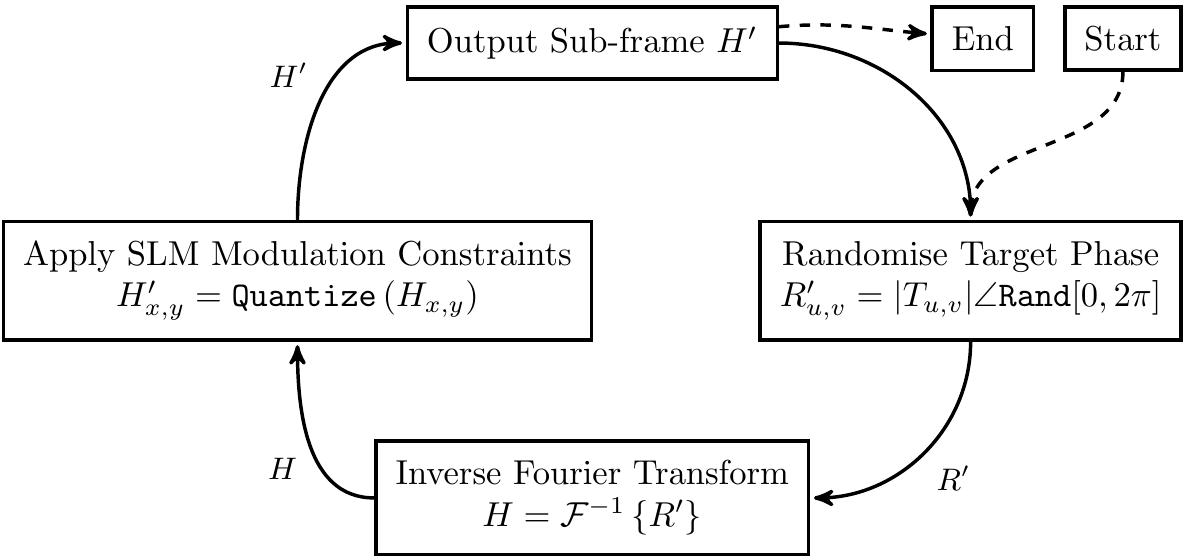}}
        {\includegraphics[width=0.48\textwidth,page=1]{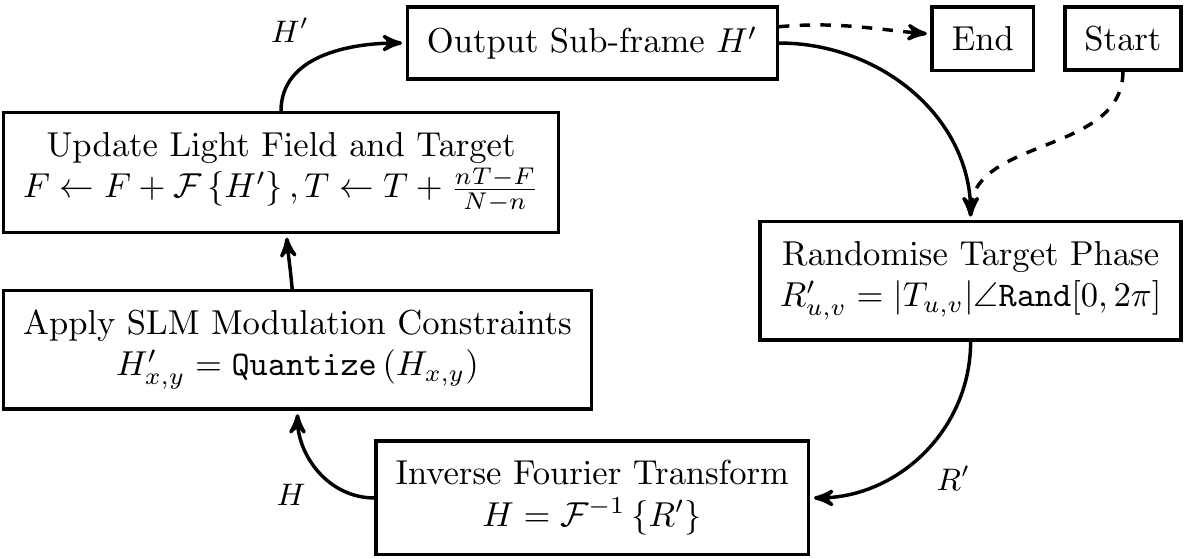}}
        \caption{Selection of time-averaging algorithms including One-Step Phase-Retrieval (left) and Adaptive One-Step Phase-Retrieval (right).}
        \label{fig:algs3}
    \end{figure}

    \subsection{Algorithm Choice}
    
\begin{figure}[tbhp]
        \centering
        {\includegraphics[width=\textwidth,page=1]{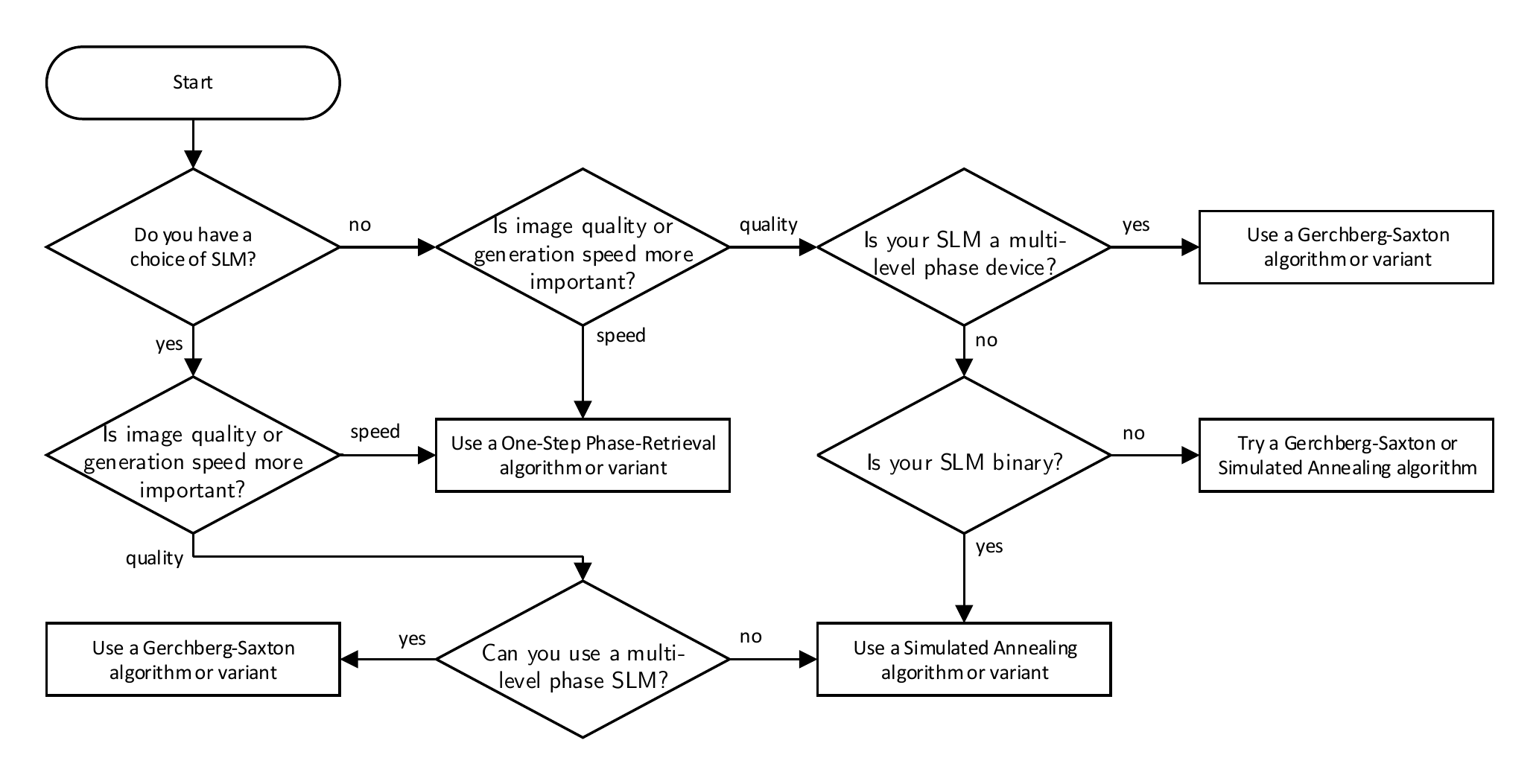}}
        \caption{Basic decision process for algorithm choice when designing a holographic system.}
        \label{fig:choice}
    \end{figure}

    The choice between the three categories of algorithms is a non-trivial decision requiring detailed knowledge of the application and SLM used. The primary considerations include the SLM modulation capabilities, the form of the target images and whether the target is phase insensitive. Figure~\ref{fig:choice} shows a simple decision flow chart.

    \section{Implementation and Structure}
    
\begin{figure}[tbhp]
        \centering
        {\includegraphics[width=1.0\textwidth]{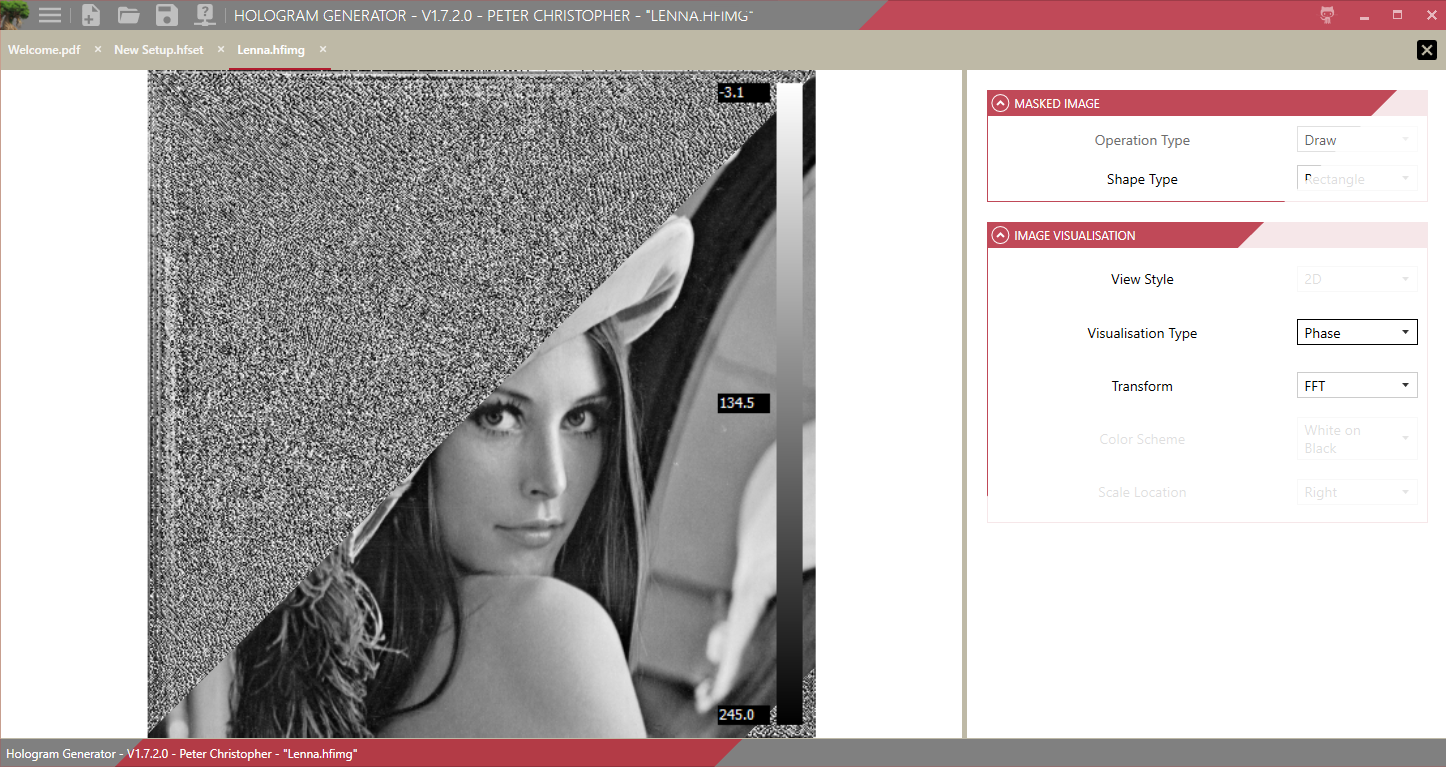}}
        \caption{Screenshot of the HoloGen application showing the target image (red) overlaid by the generated binary amplitude hologram (grey)}
        \label{fig:screenshot}
    \end{figure}

    HoloGen, Figure~\ref{fig:screenshot}, is built on a MVVMA architecture. This is a standard model-view-viewmodel (MVVM) framework commonly used in \CS{} Windows Presentation Framework (WPF) applications with an additional algorithms level written in a more traditional procedural/functional style on top of an Nvidia Cuda architecture interfaced in \CC. While the application has targeted traditional structure for ease of extension, a number of structural and implementation features deserve mention and additional detail is packaged with the source code.

    \subsection{Graphical User Interface}
    
    The Graphical User Interface (GUI) is based on the Windows presentation framework (WPF). WPF in turn uses the extensible application markup language (XAML) to define the user interface components. Like many modern GUI packages, WPF encourages \textit{binding} where elements in the \textit{view} layer are bound to properties and collections in the \textit{ViewModel} layer. This approach allows for two way data flow and removes much of the filler code found in older primarily event driven architectures such as WinForms. This approach also allows for easy runtime injection and extension, meaning that GUI portions are only loaded when required.
    
    \subsection{Reflection Parameter Hierarchy}
                
    HoloGen uses a reflection based parameter and command system. This is in contrast to the XML parameter sheet systems widely used by comparable applications. 
    Instead of the parameter types and interactions being defined in parameter sheets which are parsed at runtime, the parameter system
    is coded into the \CS{} directly. This significantly reduces the runtime overhead as well as improves the error checking available 
    at compile time. The downside is an increased architecture exposure of the parameter hierarchy.

   %
   %
   %
   %
        
    \subsection{Interop}  
        
    For fast and easy transfer of large images to the \CC{} subsystem, a three-level architecture is used. The use of managed \CC{} increases the structural complexity but allows the \CS{} application layer to be ignorant of the dynamic link library (DLL) interface. The use of the native \CC{} layer allows the use of Nvidia Thrust tools as class members.    
        

    \subsection{Fast Fourier Transforms}
    
    The majority processing factor in any holographic system is the two-dimensional Fourier transform. Our tests found that the FFT calculation or update step took over $98\%$ of the runtime for all algorithms on our system when input and output operations were excluded. As a result, any implementation is heavily dependent on the FFT library used. 
    
\begin{figure}[tbhp]
        \centering
        {\includegraphics[trim={0 0 0 0},width=0.65\linewidth,page=1]{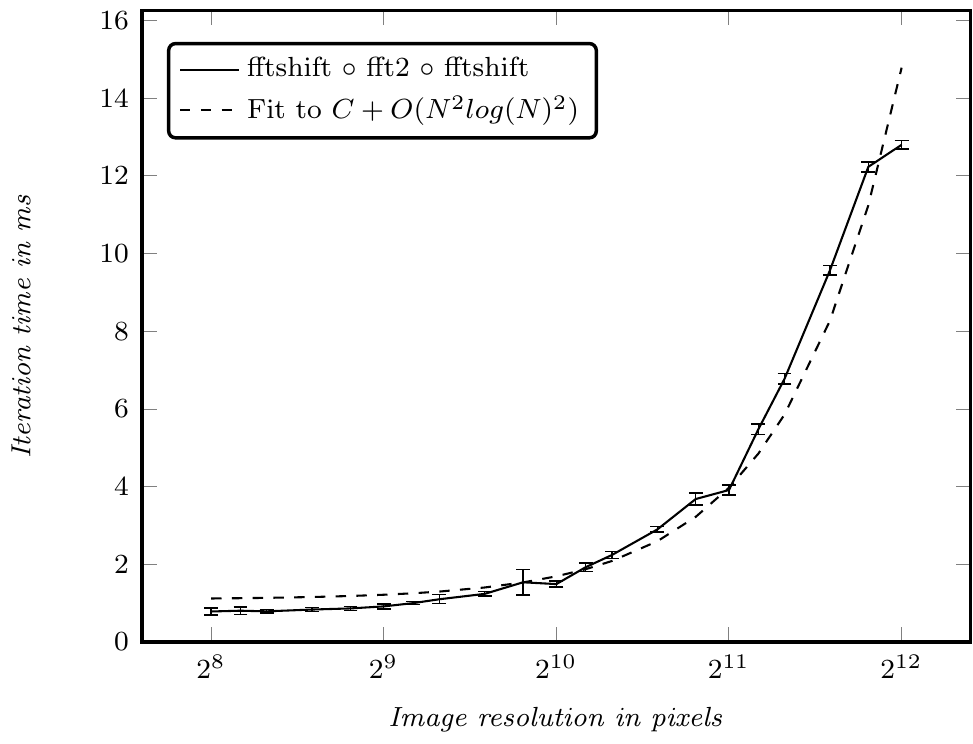}}
        \caption{Performance of cuFFT for differing square image resolutions. Error bars show the $2\sigma$ confidence interval measured from 100 independent runs of 1000 pairs of FFTs and IFFTs}
        \label{benchmarkFFT}
    \end{figure}

    HoloGen currently uses cuFFT, Nvidia's FFT implementation for their graphical processing units (GPUs), due to its high reliability and performance \cite{steinbach2017gearshifft}. A graph of the performance of cuFFT against resolution is shown in Figure~\ref{benchmarkFFT} along with the idealised $O(N^2\log{N})$ trend line.     
    
    \subsection{Floating Points}
    
    The IEEE standards define 32-bit and 64-bit floating point numbers, represented in \CC{} by \code{single} and \code{double} values \cite{IEEE7541985}. Less widely used is the 16-bit floating point \cite{IEEE7542008}. 16-bit numbers are ideal for GPU based computation and in particular holography where error is less likely to be cumulative. Real-world image formats are typically 8-bit per colour meaning that a 16-bit floating point, when scaled correctly, can more than accommodate the necessary information while significantly calculation improving performance. The scaling element is key for 16-bit operations where care must be taken to normalise all FFT operations to reduce unexpected errors and overflows.
    
    HoloGen is capable of being compiled in 16-, 32- and 64- bit versions with the application performance being approximately proportional to the reciprocal of the number of data bits. HoloGen also automatically scales every image in order to increase accuracy for low numbers of bits.

    \subsection{Templatisation}        
    
    The standard version of HoloGen tracks properties such as the illumination fields that are not necessary in some applications. By making significant use of the \CC{} template syntax, this can be tuned at compile time. This allows compile time flexibility in the required algorithm portions while still offering runtime performance.
    
    This is combined with the Nvidia runtime compilation (NVRTC) which allows users the ability to modify algorithms at runtime. This is not currently exposed in the GUI for HoloGen but is available in the application programming interface (API).
    
    \subsection{Example}
    
    The code listing in Figure~\ref{fig:listing1} demonstrates a number of these principles in action. The \code{struct} shown, \code{quantiseDiscretePhase}, exposes the \code{()} operator. The Thrust library is used to call this as shown in Figure~\ref{fig:listing2} where Thrust handles the memory management of calling the \code{quantiseDiscretePhase} operator on its arguments. Properties such as \code{FloatType} and \code{IntType} allow for changing the numerical representation at runtime while use of the \code{if constexpr} syntax from \CC{}17 allows for unwanted execution pathways to by ignored. Modern \CC{} allows for significant flexibility between runtime (\code{const}) and compile time (\code{constexpr}) constness. By changing the \code{DerivedPolicy}, it is possible to compile the application for CPU or for GPU operation.

    \begin{figure}
        \begin{lstlisting}[language=C++]
template<bool FullCircle,typename FloatType,typename IntType>
struct quantiseDiscretePhase {
private:
    constexpr float _pi=3.14159265359; constexpr float _pi2=6.28318530718;
    inline constexpr const FloatType ConstrainSLM(const FloatType& diffArg,constexpr FloatType illumArg,
        constexpr FloatType illumAbs) {
        if constexpr (!FullCircle) {
            while ((diffArg-_maxSLMArg)>_pi2) diffArg-=_pi2;
            while ((diffArg-_minSLMArg)<-_pi2) diffArg+=_pi2;
            if (diffArg > _maxSLMArg) 
                return thrust::polar<FloatType>(illumAbs,
                    illumArg+(diffArg<_wrapMaxSLMArg?_maxSLMArg:_minSLMArg));
            if (diffArg<_minSLMArg) 
                return thrust::polar<FloatType>(illumAbs,
                    illumArg+(diffArg > _wrapMinSLMArg?_minSLMArg:_maxSLMArg));
        }
        return diffArg;
    }
    const FloatType _minSLMArg; const FloatType _maxSLMArg;
    const IntType _levels; const FloatType _spac;
    const FloatType _wrapMinSLMArg; const FloatType _wrapMaxSLMArg;
public:
    quantiseDiscretePhase(const FloatType minSLMArg,const FloatType maxSLMArg,const IntType levels) :
        _minSLMArg(minSLMArg),_maxSLMArg(maxSLMArg),_levels(levels), _spac((_maxSLMArg-_minSLMArg)/(_levels-1)),
        _wrapMinSLMArg(maxSLMArg+fmod((maxSLMArg-minSLMArg),_pi2)/2.0),
        _wrapMaxSLMArg(minSLMArg-fmod((maxSLMArg-minSLMArg),_pi2)/2.0) {};
    __device__ constexpr const thrust::complex<FloatType> operator()(const thrust::complex<FloatType>& input) {
        const auto inputArg=Globals::Arg(input.real(),input.imag());
        const auto diffArg=ConstrainSLM(fmod(inputArg,_pi2),0,1);
        const auto discArg= _minSLMArg+_spac*roundf((diffArg-_minSLMArg)/_spac);
        return thrust::polar<FloatType>(1,discArg);
    }
};
    \end{lstlisting}
    \caption{HoloGen quantisation operator for nearest-neighbour quantisation for a discrete phase level SLM.}
    \label{fig:listing1}
\end{figure}

\begin{figure}
    \begin{lstlisting}[language=C++]
    thrust::transform(DerivedPolicy,input.begin(),input.end(),
        thrust::make_zip_iterator(thrust::make_tuple(IlluminationPhases->begin(),IlluminationMagnitudes->begin())),
        input.begin(), quantiseDiscretePhase<FullCircle,FloatType,IntType>(MinSLMValue, MaxSLMValue, Levels));
    \end{lstlisting}
    \caption{Calling a operator using Thrust.}
    \label{fig:listing2}
\end{figure}

    \subsection{Technical Documentation}
    
    A number of technical documents are available for more information. A detailed overview of the packages, libraries and main classes can be found on the arXiv. \cite{christopher2020structure} A guide to building, deployment and source code editing can be found in the \code{README.md} file in the repository\footnote{\url{https://gitlab.com/CMMPEOpenAccess/HoloGen/README.md}}. A Doxygen compilation of in package documentation can also be found in the repository\footnote{\url{https://gitlab.com/CMMPEOpenAccess/HoloGen/Documentation/refman.pdf}}.
                
    \section{Validation}
    
    HoloGen was developed with a research group with significant ongoing research in holographic algorithms and applications and has been incorporated in a number of experimental systems \cite{MostChangedPixel,Christopher:19,CHRISTOPHER2020124666,CHRISTOPHER2020125353,CHRISTOPHER2020125883,CHRISTOPHER2020125701,christopher2019improving,christopher2020linear,christopher2019variance,Kadis_2020}. 
    
\begin{figure}[tbhp]
        \centering
        {\includegraphics[width=1.0\textwidth]{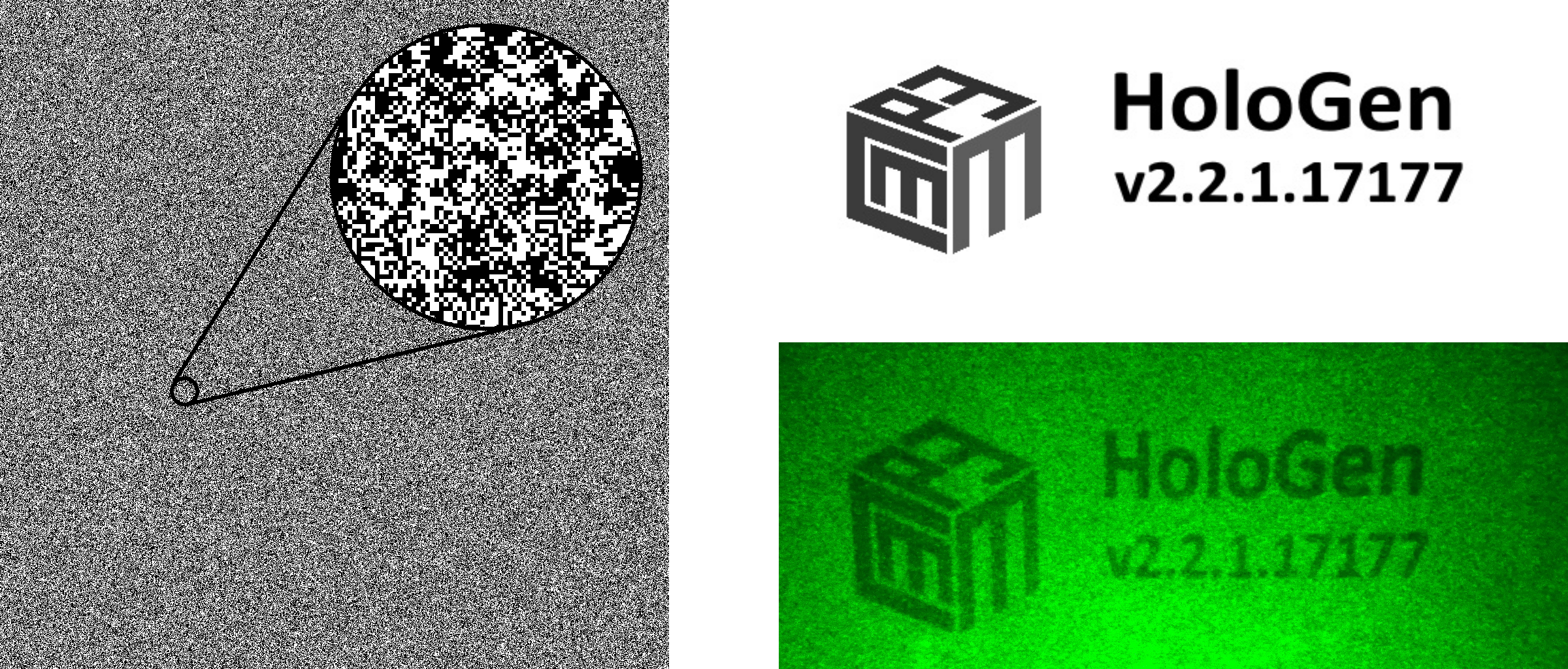}}
        \caption{Binary phase hologram generated with HoloGen including target image (top right), single binary subframe (left) and measured result (bottom right). Captured using a Canon 5D Mark III with a 24-105mm lens.}
        \label{fig:test}
    \end{figure}

    Figure~\ref{fig:test} shows a binary-phase OSPR image generated using HoloGen for a $512\times512$ pixel image showing the generated hologram (left), target image (top right) and result (bottom right). The poor reproduction quality is due to the limitations of the optical setup used. 
    
    \section{Conclusion}
    
    In this paper we have introduced a novel open-source software package for generating two-dimensional holograms suitable for fibre and wavelength multiplexing, image correction, image recognition, optical tweezing and video projection. A brief introduction to computer-generated holography was presented and the implemented algorithms introduced. The structure and features of HoloGen were also discussed before an example of HoloGen in practice was presented. 
    
    \section*{Acknowledgements}
    
    PJC acknowledges funding from the Engineering and Physical Sciences Research Council (EP/L016567/1, EP/T008369/1 and EP/V055003/1). GSDG acknowledges funding from Cancer Research UK (C47594/A21102, C55962/A24669); and a pump-priming award from the CRUK Cambridge Centre Early Detection Programme (A20976)
        
    \bibliography{references}
    
\end{document}